\documentclass{aastex}
\usepackage{amsmath}
\usepackage{latexsym,bm}
\usepackage{subfigure}
\usepackage{graphicx}
\usepackage{color}

\begin{document}

\title{Whether the Classical O-C Diagram can be used to Multi-mode Pulsators}

\author{Jia-Shu Niu}
\affil{ Institute of Theoretical Physics, Chinese Academy of Sciences, Beijing, 100190, P.R.China}
\author{Yang Wang}
\affil{School of Mathematical Sciences, Beijing Normal University,    Beijing 100875, P.R.China}
\author{Jian-Ning Fu\thanks{Send offprint request to: jnfu@bnu.edu.cn}}
\affil{Astronomy Department, Beijing Normal University,    Beijing 100875, P.R.China}
\author{Hui-Fang Xue}
\affil{Astronomy Department, Beijing Normal University,    Beijing 100875, P.R.China}

\begin{abstract}
O-C diagram is a useful technique to analyse the period changes of a pulsator by using the maximum (or minimum) value points that have been obtained from the historical data. But if an object is a double-mode or multi-mode pulsator, the extreme value points are the results of all the modes other than just the fundamental mode. We will discuss these situations and give out some criteria to judge whether O-C diagram can be used under these situations. 

\end{abstract}


\section{Introduction}
\label{sec-1}

O-C means O[bserved] minus C[alculated], it is a useful method that we always use to analyse the period change of variables. Moreover, we do not plan to discuss the details how to do O-C diagram, which would be found in \cite{Sterken2005}.

 The usage of O-C diagram in variables depends on the maximum value points of the light curves we observed, but these points are always unstable when the object is a double-mode or multi-mode pulsator. All the modes coupling together and the maximum points are the comprehensive effects of them all. Of course, we can use the other methods to analyse the period change of the pulsator for example the Fourier analysis for each data set. But in most cases, these methods need a lot of high quality data, which is always difficult to obtain. Furthermore, in the historical observational data of most variables, the maximum value points are the most credible and always the only credible parts. We should make full use of the data. As a result, we give out some criteria to decide whether the classical O-C diagram can be used to these pulsators.

In the following sections, we present a detailed discussion of how to estimate transformation of the maximum value points and give some criteria which can be compared with the observational errors and fit errors of O-C diagram. We also assume that a global pulsation analysis has already been made, and the frequencies of the pulsator have been obtained.

The organization of the paper is: Section 2 analyses the transformation of the maximum values and gives out the criteria; Section 3 applies the criteria to a specific example; Section 4 gives our conclusions.

\section{The error estimation}
\label{sec-2}
\subsection{Combination of two different sinusoidal functions}
\label{sec-2-1}

Let,
\begin{enumerate}
\item $f_{0}(t) = A_{0} \cos(\omega_{0}t + \phi_{0})$;
\item $f_{1}(t) = A_{1} \cos(\omega_{1}t + \phi_{1})$;
\item $f_{0,1}(t) = A_{0} \cos(\omega_{0}t + \phi_{0}) + A_{1} \cos(\omega_{1}t + \phi_{1})$.
\end{enumerate}

In the above equations, we choose $f_{0}$ and $f_{1}$ satisfying $\frac{A_{1} \omega_{1}}{A_{0} \omega_{0}} < 1$. In most conditions, $A_{0} > A_{1}$ and $\omega_{0} < \omega_{1}$, but this is not always the case.

The main problem can be reduced to the problem below:
let

\begin{itemize}
\item $f_{0}'(t^{m}_{0}) = 0$ and $f_{0}''(t^{m}_{0}) < 0$;
\item $f_{0,1}'(t^{m}_{0,1}) = 0$ and $f_{0,1}''(t^{m}_{0,1}) < 0$,
\end{itemize}

and then we want to get $t^{m}_{0}$ and $t^{m}_{0,1}$, which are the maximum value points of $f_{0}$ and $f_{0,1}$, respectively.

For any value of $t^{m}_{0}$, we choose the $t^{m}_{0,1}$ that is closest to $t^{m}_{0}$, and let $\delta T^{m} = \vert t^{m}_{0}-t^{m}_{0,1} \vert$. And we are here to want to estimate $\Delta T^{m} = max\{\delta T^{m} \}$.

\subsubsection{$\omega_{1} = n \omega_{0}, n \in \mathbb{N}^{+}$}
\label{sec-2-1-1}

In this case, $f_{0,1}(t)$ is the combination of the harmonics of $\omega_{0}$, which do not influence the distance between two maximum values. Thus:  $\delta T^{m} = 0$.

\subsubsection{$\omega_{1} = \alpha \omega_{0}, \alpha \in \mathbb{R}^{+} \setminus \mathbb{N^{+}}$}
\label{sec-2-1-2}

The primary problem can be reduced to the problem below:
\begin{enumerate}
\item $f_{0}(t) = A_{0} \cos(\omega_{0}t)$;
\item $f_{1}(t) = A_{1} \cos(\omega_{1}t + \phi_{1})$;
\item $f_{0,1}(t) = A_{0} \cos(\omega_{0}t) + A_{1} \cos(\omega_{1}t + \phi_{1})$;
\end{enumerate}

the left are the same as before. All the phase effects attribute to $\phi_{1}, \phi_{1} \in \mathbb{R}$. Note that when $\phi_{1}$ is a free parameter, all the other cycles can be represented by the cycle near $t = 0$ if we choose a proper $\phi_{1}$.

We now focus on the maximum value near $t = 0$. We know that $t^{m}_{0} = 0$, and let us estimate $t^{m}_{0,1}$.

Let $f_{0,1}'(t) = 0$, leading to:
\begin{equation}
A_{0}\omega_{0} \sin(\omega_{0}t) + A_{1}\omega_{1} \sin(\omega_{1}t + \phi_{1}) = 0.
\end{equation}

Then,
\begin{equation}
\sin(\omega_{0}t) = -\frac{A_{1}\omega_{1}}{A_{0}\omega_{0}}sin(\omega_{1}t + \phi_{1}) \le \frac{A_{1}\omega_{1}}{A_{0}\omega_{0}}.
\end{equation}

When the equal sign is satisfied, $A_{1}sin(\omega_{1}t + \phi_{1})$ reaches its maximum value, and  $\sin(\omega_{0}t) = \frac{A_{1}\omega_{1}}{A_{0}\omega_{0}}$. Since sinusoidal function is monotonous in $[-\pi,\pi]$, $t^{m}_{0,1}$ reaches its largest value now. At the same time, $t^{m}_{0} = 0$, therefore we can get:
\begin{equation}
\Delta T^{m} = \max\{\delta T^{m}\} = \frac{1}{\omega_{0}} \arcsin(\frac{A_{1}\omega_{1}}{A_{0}\omega_{0}}).
\end{equation}

\subsection{The transformation of two adjacent maximum values}
\label{sec-2-2}

Because all the values of $t^{m}_{0,1}$ are around the corresponding values of $t^{m}_{0}$, and if $\omega_{1} / \omega_{0}$ is a irrational number, then all the values in $[0, \Delta T^{m}]$ will be reached in some cycles (Of course, $\omega_{0} / \omega_{1}$ ate always in this case). Averagely speaking, we have

\begin{equation}
\lim_{n\to\infty} \frac{\sum_{n} (\vert t^{m}_{0,1}(n) - t^{m}_{0,1}(n+1) \vert )}{n} = \lim_{n\to\infty} \frac{\sum_{n} (\vert t^{m}_{0}(n) - t^{m}_{0}(n+1) \vert)}{n} = \frac{2 \pi}{\omega_{0}}.
\end{equation} 

In the equation, $t^{m}_{0,1}(n)$ and $t^{m}_{0}(n)$ imply the n-th maximum values of $f_{0,1}$  and $f_{0}$ from $t = 0$.

When we use O-C method practically, the transformation of two adjacent maximum values ($\delta t^{m} = \vert{\vert{t^{m}_{0,1}(n)-t^{m}_{0,1}(n+1)\vert - \frac{2\pi}{\omega_{0}}\vert}}$) are more important than the absolute transformation ($\Delta T^{m}$) relative to $t^{m}_{0}$. In other words, the upper limit of distance between two adjacent maximum values $\Delta t^{m} = \max\{\delta t^{m}\}$ should be estimated.

In this case, let:
\begin{itemize}
\item $f_{0}(t) = A_{0} \cos(\omega_{0}t)$, $f_{0}'(t) = -A_{0}\omega_{0} \sin(\omega_{0}t)$;
\item $f_{1}(t) = A_{1} \cos(\omega_{1}t + \phi_{1})$, $f_{1}'(t) = -A_{1}\omega_{1} \sin(\omega_{1}t + \phi_{1})$.
\end{itemize}

In general conditions, $\phi_{0}$ and $\phi_{1}$ are not important if we only want to get a upper limit of $\delta t^{m}$.

Let us calculate the next period after $t=0$:
\begin{itemize}
\item $f_{0}'(t+T_{0}) = -A_{0}\omega_{0} \sin(\omega_{0}(t + \frac{2\pi}{\omega_{0}}))$;
\item $f_{1}'(t+T_{1}) = -A_{1}\omega_{1} \sin(\omega_{1}(t + \frac{2\pi}{\omega_{1}}) + \phi_{1})$.
\end{itemize}

For $\omega_{0} < \omega_{1}$, $\frac{2\pi}{\omega_{0}} > \frac{2\pi}{\omega_{1}}$, let $\Delta \phi = \frac{2\pi}{\omega_{0}} - \frac{2\pi}{\omega_{1}}$.

Then,
\begin{itemize}
\item $A_{0}\omega_{0} \sin(\omega_{0}t) = -A_{1}\omega_{1} \sin(\omega_{1}(t + \Delta \phi) + \phi_{1})$.
\end{itemize}
This can be reduced to the condition in section 2.1, and the solution of this equation near $t=0$ stands for the $\delta t^{m}$ between the first and the second maximum values of $f_{0,1}$.

Obviously, $\delta t^{m} \le \Delta T^{m}$ and $\delta t^{m}$ changes along with $\Delta \phi$ periodically.

Note that $\Delta T^{m} = \frac{1}{\omega_{0}} \arcsin(\frac{A_{1}\omega_{1}}{A_{0}\omega_{0}})$, and $\Delta \phi$ depends on the $\omega_{0}$ and $\omega_{1}$, we push out a estimation of $\delta t^{m}$ (For every two adjacent cycles, the phase differences would be $\Delta \phi$.):
\begin{equation}
\delta t^{m} = \Delta T^{m} \sin(\frac{\pi}{\Delta T^{m}} \Delta \phi) \le \Delta T_{m}.
\end{equation}

As a result, $\Delta t^{m}$ could be given as:
\begin{equation}
\Delta t^{m} = \max\{\delta t^{m}\} = \Delta T^{m} \sin(\frac{\pi}{T^{m}} \Delta \phi - \frac{\pi}{2}) - \Delta T^{m} \sin(-\frac{\pi}{2}) = \Delta T^{m} \sin(\frac{\pi}{T^{m}} \Delta \phi - \frac{\pi}{2}) + \Delta T^{m}.
\end{equation}
\subsection{Combination of three different sinusoidal functions}
\label{sec-2-3}
\subsubsection{Harmonics}
\label{sec-2-3-1}

    For harmonic terms, the maximum value points of $\omega_{0}$ would not be transformed.
\subsubsection{Three absolute modes}
\label{sec-2-3-2}
\begin{itemize}

\item Estimation of $\Delta T^{m}$\\
\label{sec-2-3-2-1}%
A proper estimation can be given as follows, if we use the same method as before.
\begin{equation}
A_{0}\omega_{0} \sin(\omega_{0}t) = -A_{1}\omega_{1} \sin(\omega_{1}t + \phi_{1}) - A_{2}\omega_{2} \sin(\omega_{2}t + \phi_{2}) \le \vert A_{1}\omega_{1} +  A_{2}\omega_{2} \vert
\end{equation}
If 
\begin{equation}
\vert \frac{A_{1}\omega_{1} +  A_{2}\omega_{2}}{A_{0}\omega_{0}} \vert \le 1,
\end{equation}
then we can get
\begin{equation}
\Delta T^{m} = \frac{1}{\omega_{0}} \arcsin(\frac{A_{1}\omega_{1} + A_{2}\omega_{2}}{A_{0}\omega_{0}}).
\end{equation}

\item Estimation of $\Delta t^{m}$\\
\label{sec-2-3-2-2}%
If we take the total effects as the additional effects of $\omega_{1}$ and $\omega_{2}$ independently, then we can use their linear superposition as the estimation of $\Delta t^{m}$.
Consequently, the estimation should be
\begin{equation}
\Delta t^{m}  = \Delta T^{m}_{1} \sin(\frac{\pi}{T^{m}_{1}} \Delta \phi_{1} - \frac{\pi}{2}) + \Delta T^{m}_{1} + \Delta T^{m}_{2} \sin(\frac{\pi}{T^{m}_{2}} \Delta \phi_{2} - \frac{\pi}{2}) + \Delta T^{m}_{2}.
\end{equation}
$\Delta T^{m}_{1}$, $\Delta T^{m}_{2}$, $\Delta \phi_{1}$ and $\Delta \phi_{2}$ are defined similarly.

\end{itemize} 
\subsection{Generalization}
\label{sec-2-4}
\subsubsection{Harmonics}
\label{sec-2-4-1}

In practical usage, all the harmonic terms of $\omega_{p}$ ($p \in \mathbb{N}$) could be estimated totally as one trigonometric function for convenience. We use one sine function to estimate the total effects of the all the terms of $\omega_{p}$.
\begin{equation}
f_{p}(t) = A_{p} \cos(\omega_{p}t + \phi_{p}) \approx \bar f_{p} (t) = \sum_{q} A_{p,q} \cos(q\omega_{p}t + \phi_{p,q}). q \in \mathbb{N}, q > 1
\end{equation}

In this equation, $A_{p} = \max_{t \in mathbb{R}}\{\bar f_{p} (t)\}$ and $\phi_{p}$ ensures $f_{p}(t) = \bar f_{p} (t)$ when $t = t^{m}_{p}$ ($t^{m}_{p}$ is the maximum value points of $\bar f_{p} (t)$).

Because we use this function to estimate the maximum values, and in most cases, $A_{p,q} / A_{p,q-1} > 5$, $\bar f_{p} (- \frac{\phi_{p,1}}{\omega_{p}})$ would be a good approximation of $A_{p}$ and $\phi_{p,1}$ would be a good approximation of $\phi_{p}$.

As a consequent, we choose
\begin{equation}
f_{p}(t) = A_{p} \cos(\omega_{p}t + \phi_{p}).
\end{equation}
to reprensent the effects of $\bar f_{p} (t)$ (sometimes we can also choose $A_{p} = \sum_{q} A_{p,q}$ as the approximation of $A_{p}$).
\subsubsection{Multi-mode combinations}
\label{sec-2-4-2}

We just consider a $p-mode$ ($p > 1$) pulsator as an example. Using the method above, we obtain the generalization as follows.

If 
\begin{equation}
\vert \sum \frac{A_{p}\omega_{p}}{A_{0}\omega_{0}} \vert \le 1,
\end{equation}
then we can give another estimation:
\begin{equation}
\Delta T^{m} = \frac{1}{\omega_{0}} \arcsin(\sum_{p} \frac{A_{p}\omega_{p}}{A_{0}\omega_{0}}).
\end{equation}

The estimation of transformation of adjacent maximum values should be:
\begin{equation}
\Delta t^{m}  = \sum_{p}(\Delta T^{m}_{p} \sin(\frac{\pi}{\Delta T^{m}_{p}} \Delta \phi_{p} - \frac{\pi}{2}) + \Delta T^{m}_{p}).
\end{equation}

In the equation above, $\Delta T^{m}_{p} = \frac{1}{\omega_{0}} \arcsin(\frac{A_{p}\omega_{p}}{A_{0}\omega_{0}})$ and $\Delta \phi_{p} = \frac{2\pi}{\omega_{0}} - \frac{2\pi}{\omega_{p}}$.

\section{An example (AE UMa)}
\label{sec-3}

AE UMa is a SX Phe star which has two modes of pulsation. Because of its double-mode pulsation, it is doubtful that whether the classical O-C method can be used to analyse the period change of its fundamental mode. The key is that the maximum values are the results of two mode together, the first overtone and the beat mode cause the transformation of the maximum values. Now, let's estimate the upper limit of the transformation by using the criteria introduced above.
\subsection{Estimation of $\Delta T^{m}$ and $\Delta t^{m}$}
\label{sec-3-1}

First, using the data from \cite{Niu2013}, Table 3, we collect each pulsation terms $\omega_{0}$ and $\omega_{1}$. 

Let's deal with the combinations of $\omega_{0}$ and $\omega_{1}$ ($A_{m \omega_{0}, n \omega_{1}} \cos((m \omega_{0} + n \omega_{1}) t + \phi_{m \omega_{0}, n \omega_{1}})$, $m, n \in \mathbb{Z}$). 

All these terms can be transform to the form
\begin{multline}
A_{m \omega_{0}, n \omega_{1}} \cos((m \omega_{0} + n \omega_{1}) t + \phi_{m \omega_{0}, n \omega_{1}}) \\
= A_{m \omega_{0}, n \omega_{1}} \sin(\omega_{1}t + \varphi_{m \omega_{0}, n \omega_{1}}) \cos(\omega_{0}t + \varphi_{m \omega_{0}, n \omega_{1}}) +\\
 A_{m \omega_{0}, n \omega_{1}} \sin(\omega_{0}t + \varphi_{m \omega_{0}, n \omega_{1}}) \cos(\omega_{1}t + \varphi_{m \omega_{0}, n \omega_{1}}).
\end{multline}
In the equation, $\varphi_{m \omega_{0}, n \omega_{1}} = \frac{1}{2} (\phi_{m \omega_{0}, n \omega_{1}} -  \frac{\pi}{2})$.

And we can add the term $A_{m \omega_{0}, n \omega_{1}} \sin(\omega_{1}t + \varphi_{m \omega_{0}, n \omega_{1}}) \cos(\omega_{0}t + \varphi_{m \omega_{0}, n \omega_{1}})$ to $\bar f_{0} (t)$ and the term $A_{m \omega_{0}, n \omega_{1}} \sin(\omega_{0}t + \varphi_{m \omega_{0}, n \omega_{1}}) \cos(\omega_{1}t + \varphi_{m \omega_{0}, n \omega_{1}})$ to $\bar f_{1} (t)$.

 In addition, we only choose the frequencies whose amplitude larger than $10  mmag$, since the smaller ones are considered to influence the result slightly.

The reduced $f_{0}$ and $f_{1}$ can be obtained by $A_{p} = f_{p} (- \frac{\phi_{p}}{\omega_{p}})$ and $\phi_{p,1}$. According to the calculation we introduced: $\Delta T^{m} = 0.002728$, $\Delta t^{m} = 0.001504$. This result implies that the upper limit of the transformation between adjacent maximum values is in the same level with the observational errors and the fit error from O-C diagram, which have the value of $0.002083$ and $0.00244$ respectively.

As a result, the classical O-C method can be used to determine the period changes in this situation, but just for the fundamental mode.

\section{Conclusion}
\label{sec-4}

In this paper, we give out the criteria which can tell one whether the classical O-C method could be used to analyse the maximum values observed from a multi-mode pulsator. According to comparing the upper limit estimation of the maximum value transformation ($\Delta T^{m}$, $\Delta t^{m}$) and the observational errors or fit errors of O-C diagram, we can make a decision whether the O-C method can be used effectively. If $\Delta t^{m} < \sigma$, we think it is meaningful to use the O-C diagram to do analysis.


\end{document}